# First-principles phonon calculations of thermal expansion path of $Fe_2Mo$ Laves phase

**Dmitry Vasilyev**

Baikov Institute of Metallurgy and Materials Science of RAS, 119334, Moscow, Leninsky Prospekt 49, Russia

dvasilyev@imet.ac.ru ; vasilyev-d@yandex.ru

**Abstract**. Precipitation of the topologically close-packed $Fe_2Mo$ Laves phase at the interface between the nuclear fuel and the fuel element cladding can significantly weaken the strength characteristics of the cladding and fuel. Despite the importance of designing materials for the cladding of fuel rods, the thermodynamic properties and the trajectory of the thermal expansion path of the $Fe_2Mo$ remain poorly understood. The thermodynamic properties of the $Fe_2Mo$ have been studied using the finite-temperature quantum mechanical calculations within the frame of the density functional theory (DFT) under the quasiharmonic approximation. The vibrational contribution to the free energy was obtained using phonon calculations. The thermal expansion path of $Fe_2Mo$ was predicted by comparing between free energies calculated in different directions. A path with the least energy was chosen as the trajectory of thermal expansion. The obtained result was compared with the path calculated in previous theoretical work used the Debye – Grüneisen approach and accounted magnetic subsystem to calculate the vibrational and magnetic contributions to the free energy. This comparison reveals that these two approaches are in good agreement with each other. The work shows that the $Fe_2Mo$ possesses a non-isotropic thermal expansion. The heat capacity and volumetric expansion at constant pressure are modelled. The calculated results analyzed and are in satisfactory agreement with the experimental data. Obtained results can be useful for further design of fuel element cladding materials intended for generation IV reactors, the operating temperature of which should be above 873 K.

## 1. Introduction

The Laves phase $Fe_2Mo$ is an intermetallic compound that can be precipitated from solid solutions as a result of product operation at high temperatures and under irradiation conditions. On the one hand, the $Fe_2Mo$ is considered to be a useful hardening phase, since its precipitation from the solid solution of structural steels can improve toughness, creep and oxidation resistance [1]. On the other hand, a loss of toughness of ferritic-martensitic steels is directly related to the precipitation of the $Fe_2Mo$ [2], and the precipitation of $Fe_2Mo$ at the interface between the fuel element cladding and nuclear fuel can cause deterioration of their mechanical properties [3].

$Fe_2Mo$ belongs to the Fe-Mo system, which is one of the basic binary phase diagrams used to design ferritic steels.

Ferritic steels are promising candidates for creating materials that are planned to be used in the production of fuel rods and vessel claddings for nuclear reactors of the IV generation, and the first wall in fusion reactors [4], this is due to their good resistance to swelling and creep caused by irradiation, good thermal conductivity and low coefficient of thermal expansion [5, 6].

The precipitation of $Fe_2Mo$ leads to the depletion of the steel matrix in such strengthening elements as Mo or W [7]. For example, a large precipitation of $Fe_2Mo$ in the HT9 matrix leads to a deterioration in the mechanical properties of the fuel element cladding [3].

The lattice parameters of $Fe_2Mo$ were theoretically calculated in [8, 9, 10, 11] using Density Functional Theory (DFT) at T = 0 K. Experimental values of $Fe_2Mo$ parameters obtained at 1073 K were reported in [10, 12, 13].

As far as is known, the thermodynamic properties of $Fe_2Mo$ remain poorly understood. The thermal expansion path of $Fe_2Mo$ was calculated in [14] using first-principles calculations with applying the quasi-harmonic Debye–Grüneisen (QDG) approximation and taking into account the magnetic entropy to calculate the vibrational and magnetic contributions to the free energy.

In the presented work, calculations of the free energies of $Fe_2Mo$ carried out for different directions of thermal expansions were performed using DFT-based quasi-harmonic phonon approach. Comparison of the calculation results obtained in this work with the results reported in [14], where QDG method was used, shows that these results are in good agreement with each other.

Finally, the thermal expansion $V(T)$, heat capacity $C_p(T)$ of $Fe_2Mo$ were calculated along the thermal expansion path. The results can be used to construct Gibbs potentials for phase diagram calculations, as well as to obtain properties whose experimental evaluation can be difficult.

## 2. Theory and Methods

Under the quasi-harmonic approximation, the Helmholtz free energy $F(V,T)$ takes the form according to [15-17]

$$F(V,T) = E_{tot}(V) + F_{el}(V,T) + F_{vib}(V,T) + F_{mag}(V,T) - TS_{conf} \quad (1)$$

where $E_{tot}(V)$ is the static total energy at T = 0 K. $F_{el}(V,T)$, $F_{vib}(V,T)$, $F_{mag}(V,T)$ are the electronic, vibrational and magnetic free energies contributions. $S_{conf}$ is the ideal configurational entropy.

### 2.1. Electronic energy

The contribution from the thermal electron excitation $F_{el}(V,T)$ to the free energy was computed as expressed in [18]

$$F_{el}(V,T) = E_{el}(V,T) - TS_{el}(V,T), \quad (2)$$

with $E_{el}$ is given by [19, 20]

$$E_{el}(T,V) = N_A \int_{-\infty}^{\infty} n(\varepsilon,V) f(\varepsilon,T)\varepsilon d\varepsilon - N_A \int_{-\infty}^{\varepsilon_F} n(\varepsilon,V)\varepsilon d\varepsilon \quad (3)$$

where $n(\varepsilon,V)$ is the total electronic density of states (e-DOS), $f(\varepsilon,T)$ is the Fermi-Dirac distribution and $N_A$ is Avogadro constant. Where the electronic entropy $S_{el}$ takes the form

$$S_{el}(T,V) = N_A \int_{-\infty}^{\infty} n(\varepsilon,V)\big(f(\varepsilon,T)\ln f(\varepsilon,T) + (1 - f(\varepsilon,T))\ln(1 - f(\varepsilon,T))\big)d\varepsilon \quad (4)$$

### 2.2. Vibrational energy

The vibration free energy $F_{vib}(V,T)$ was calculated from phonon density of states (PDOS) as formulated in [21, 22]

$$F_{vib}(V,T) = k_B T \int_0^{\infty} \ln\left[2 \cdot sinh\frac{\hbar\omega}{2k_B T}\right] g(\omega,V)d\omega \quad (5)$$

where $\hbar$ and $k_B$ are the reduced Planck and Boltzmann constants respectively, $g(\omega,V)$ represents PDOS as a function of phonon frequency $\omega$ calculated at volume $V$.

## 2.3. Magnetic energy

In order to take into account, the magnetic phase transition of $Fe_2Mo$ Laves phase the Hillert - Jarl model described in [24] was used. According to this model the magnetic energy is expressed as

$$F'_{mag}(T) = RTLn(\beta + 1)f(\tau) \qquad (6)$$

where $\tau = T/T_c$; $T_c$ and $\beta$ are the Curie temperature and the average magnetic moment of a compound, the $f(\tau)$ is defined in [24].

Due to the fact that the magnetic internal energy is already taken into account in the DFT calculations of the total energy, the magnetic energy $F_{mag}(V,T)$ contribution to the free energy of the compound was calculated using the following expression

$$F_{mag}(V,T) = F'_{mag}(T) - F'_{mag}(0K) \qquad (7)$$

The configurational entropy $S_{conf}$ was treated by the mean-field approximation

$$S_{conf} = N_A k_B \sum_{i=1}^{n} c_i \ln c_i \qquad (8)$$

where $c_i$ is atomic concentrations.

## 2.4. First-principles calculations

The calculations were carried out using the FP-LAPW method, as implemented in the WIEN2k code [25]. The exchange–correlation interaction was treated using the generalized gradient approximation (GGA) by Perdew, Burke, and Ernzerhof (PBE) potential [26]. The first irreducible Brillouin zones of the unit cells were sampled by a 13 x 13 x 6 $k$-point mesh by employing the Monkhorst-Pack scheme [27]. The energy convergence criterion of electronic self-consistency was set to $10^{-8}$ eV/atom.

The vibrational energy was calculated from phonon spectra, which obtained through lattice dynamical simulations performed using Phonopy package [23], where a method of small atom displacements is used to calculate forces affecting the atoms. The size of a $Fe_2Mo$ supercell used in the present calculation is $2 \times 2 \times 1$. To model the Brillouin zone for calculating the forces, the $4 \times 4 \times 4$ Monkhorst–Pack mesh [27] was used.

## 3. Results and discussion.

### 3.1. Ground state properties

#### 3.1.1. Phase stability of Fe$_2$Mo

The Fe$_2$Mo Laves phase is a stoichiometric compound with the C14 structure type and the space group *P6$_3$/mmc* with No. 194 and with the *hP12* Pearson symbol. The Fe$_2$Mo lattice is schematically shown in Figure 1.

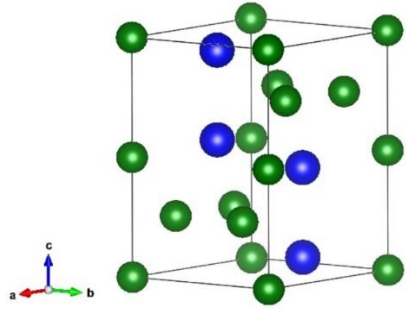

**Figure 1.** Scheme of the crystal lattice of Fe$_2$Mo Laves phase, green circles represent Fe atoms and blue circles represent Mo atoms.

All computations in this work were performed using spin-polarized calculations. The lattice parameters of Fe$_2$Mo were obtained using structural relaxation procedure. The calculated lattice parameters of Fe$_2$Mo are given in Table 1 and Figure 2, the experimental values of parameters taken in [10, 12, 13] and theoretical data [8-10, 14] are shown here too. The lattice parameters of Fe$_2$Mo calculated in this work using DFT at T = 0 K are $a$ = 8.829 and $c$ = 14.693 bohr, which is very close to the result obtained in [8] with $a$ = 8.823 and $c$ = 14.727 bohr.

**Table 1**
Lattice parameters of Fe$_2$Mo obtained in this work in comparison with the theoretical and experimental data reported in other works.

|  | $a$ (bohr) | $c$ (bohr) | $c/a$ | $V$ (bohr$^3$) |
|---|---|---|---|---|
| This work | 8.829 | 14.693 | 1.664 | 991.822 |
| Exp. [10] | 8.967 | 14.615 | 1.630 | 1017.664 |
| Exp. [13] | 8.938 | 14.589 | 1.632 | 1009.409 |
| Exp. [12] [a] | 8.940 | 14.679 | 1.642 | 1016.115 |
| This work [b] | 8.982 | 14.585 | 1.624 | 1018.986 |
| Calc. [14] [c] | 8.995 | 14.577 | 1.621 | 1021.460 |
| Calc. [8] | 8.823 | 14.727 | 1.669 | 992.842 |
| Calc. [9] | 8.804 | 14.632 | 1.662 | 982.251 |
| Calc. [10] | 8.848 | 14.286 | 1.615 | 968.528 |

[a] Obtained at T = 1073 K after 1500 hours of thermal treatment.
[b] Calculated at T = 1073K in this work using DFT-based phonon calculation.
[c] Calculated in [14] at T = 1073 K using QDG approach and magnetic entropy.

*3.1.2 Calculation outline*

In order to investigate thermal expansion paths of $Fe_2Mo$ Laves phase the numbers of possible thermal expansion trajectories were considered. Four selected trajectories (path) of $Fe_2Mo$ thermal expansion are shown in Figure 2, as an example. All these paths intersect in the point with the ($a_0$, $c_0$) coordinates, these are the equilibrium lattice parameters of $Fe_2Mo$ calculated at ground state, which listed in Table 1. Along the *n0* path the $c/a = 1.664$ ratio remains constant, this is the trajectory of isotropic thermal expansion. The *n1* path is passing through the experimental data [12] obtained at T = 1073 K after 1500 hours of annealing treatment, which shown in Figure 2 by the blue triangle. The trajectory *n2* passes through the coordinates of the lattice parameters obtained experimentally in [13], which are shown by the red triangle. The free energies calculated by (1) were obtained along each *ni* path and compared between themselves. Then, the trajectory with the least free energy was chosen as the path of thermal expansion of $Fe_2Mo$.

This calculation scheme is similar to a search of thermal expansion path (STEP) outlined in [14], but in this work the vibrational free energy was obtained using phonon dispersion curves calculated along each *ni* path. The main motivation to carry out the present work was to compare the obtained results with results calculated in [14], where the Debye–Grüneisen approach and magnetic entropy were used.

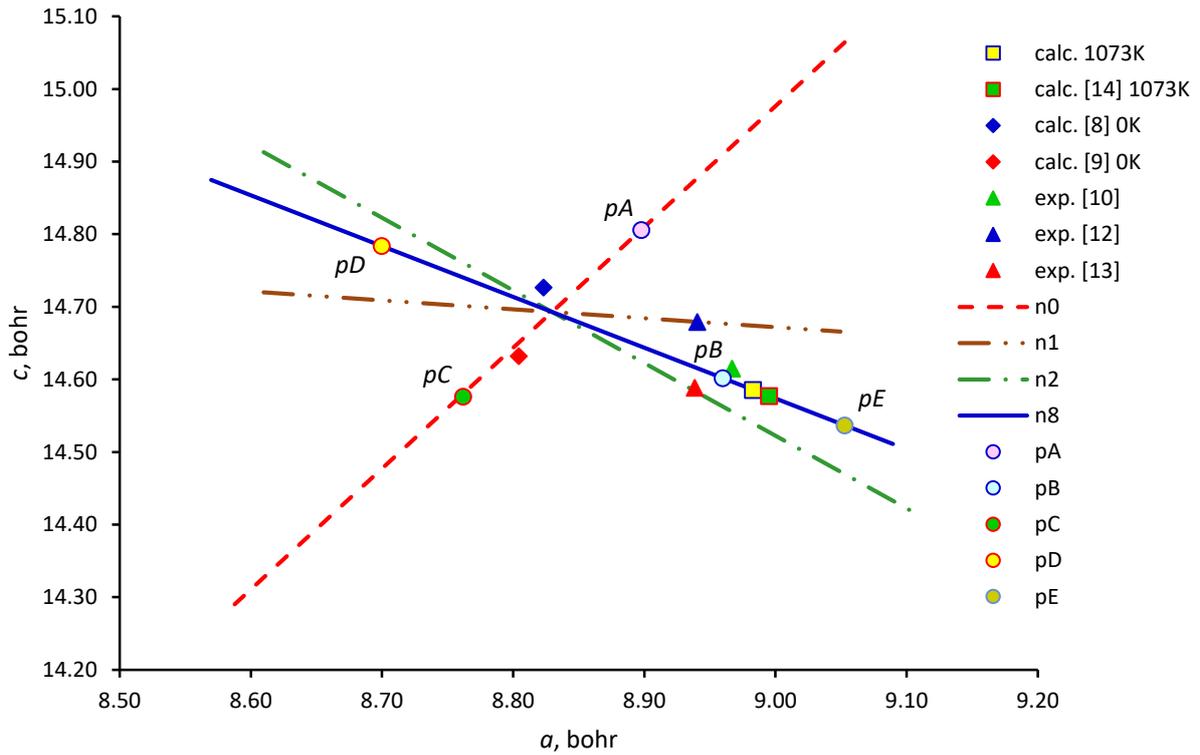

**Figure 2**. The outline of calculations. The intersection point of the *n0*, *n1*, *n2*, and *n8* paths corresponds to the lattice parameters ($a_0$, $c_0$) of $Fe_2Mo$ calculated in this work using DFT at the ground state. The theoretical data of lattice parameters calculated in [8, 9] with DFT at T = 0 K are shown by blue and red diamonds. The experimental parameters obtained at T = 1073 K in [10, 12, 13] are shown by the blue, green and red triangles. The lattice parameters of $Fe_2Mo$ calculated in this work using DFT-based phonon method and obtained in [14] at T = 1073 K by applying the quasi-harmonic Debye - Grüneisen approach and magnetic entropy are shown by the yellow and the green squares, respectively. The description of the *pA*, *pB*, *pC*, *pD* and *pE* points is in the text.

### 3.1.3 Calculations of electronic and phonon density of states

The vibrational free energies $F_{vib}(T)$ calculated by (5) make the main contribution to the Helmholtz free energy (1). The phonon dispersion curves and the corresponding phonon density of states (PDOS), shown in Figure 3, were calculated at selected points on the coordinate plane (a, c) presented in Figure 2. Figure 3 (f) plots the results obtained at the point with the lattice parameters ($a_0$, $c_0$) of $Fe_2Mo$ calculated in this work which associated with the T = 0 K. Figure 3 (a) plots the results obtained at the point $pA$ which lies on the $n0$ path, as shown in Figure 2. The other results of the phonon calculations obtained at the $pB$, $pC$ and $pD$ points are shown in Figure 3 (b), Figure 3 (c) and Figure 3 (d), respectively. The positive phonon frequencies reveal the dynamical stability of the $Fe_2Mo$ with these lattice parameters.

While, the phonon dispersions and density of states of $Fe_2Mo$ calculated at point $pE$ which is on the $n8$ path, as shown in Figure 3 (e), display extensive imaginary phonon modes, which means that at the $pE$ the $Fe_2Mo$ becomes unstable if the lattice parameters reach these values at heating. The calculation predicts that point $pE$ corresponds to T = 1291 K.

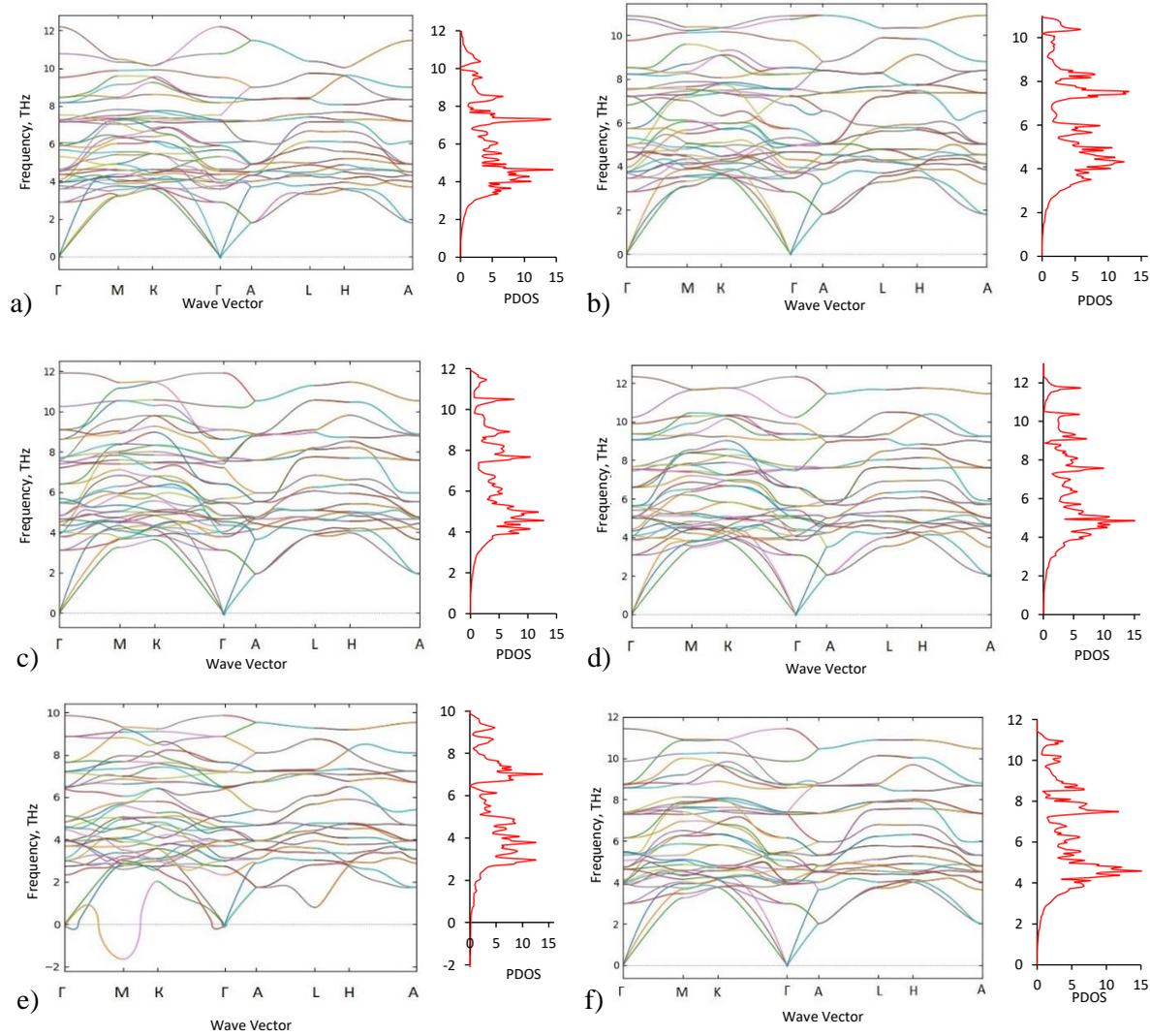

**Figure 3**. The phonon dispersion curves and the phonon density of states (PDOS) for $Fe_2Mo$ Laves phases calculated at a) point $pA$; b) point $pB$; c) point $pC$; d) point $pD$; e) point $pE$; f) point ($a_0$, $c_0$).

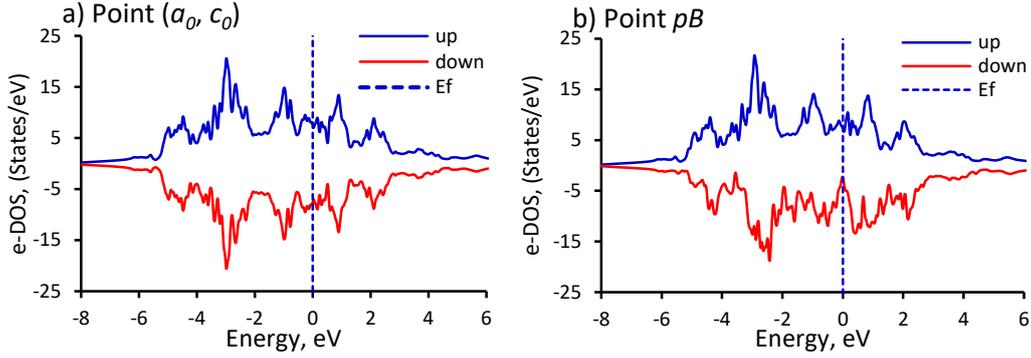

**Figure 4**. Total electron density of states (e-DOS) for $Fe_2Mo$ Laves phases calculated a) at point ($a_0$, $c_0$) and b) at point $pB$.

The electron energies (2) were calculated along all $ni$ directions, shown in Figure 2, through electron densities, according to (3) and (4). An example of such calculations, that is the total electronic density of states (e-DOS) of $Fe_2Mo$ calculated at points ($a_0$, $c_0$) and $pB$ are sown in Figure 4 (a) and Figure 4 (b), respectively.

### 3.2. Compound stability at finite temperatures

#### 3.2.1. Calculations of free energies

All the energy contributions to the Helmholtz free energy of $Fe_2Mo$, namely, total, electronic (2), vibrational (5) and magnetic (7) energies were calculated along each $ni$ path selected in Figure 2. Then, for each $ni$ path the free energy $F(V,T)$ of $Fe_2Mo$ was obtained according (1) as a sum of these energy parts. Figure 5 demonstrates an example of the free energy $F(V,T)$ obtained along $n8$ path, calculated at different temperatures and volumes. The total energy calculated at T = 0 K and equilibrium volumes $V_0$, which were obtained at the minimum of thermodynamic functions, are shown in Figure 5 by the red dashed and blue dotted lines, respectively. The thermal expansion can be seen as an increase in the equilibrium volume shown in Figure 5.

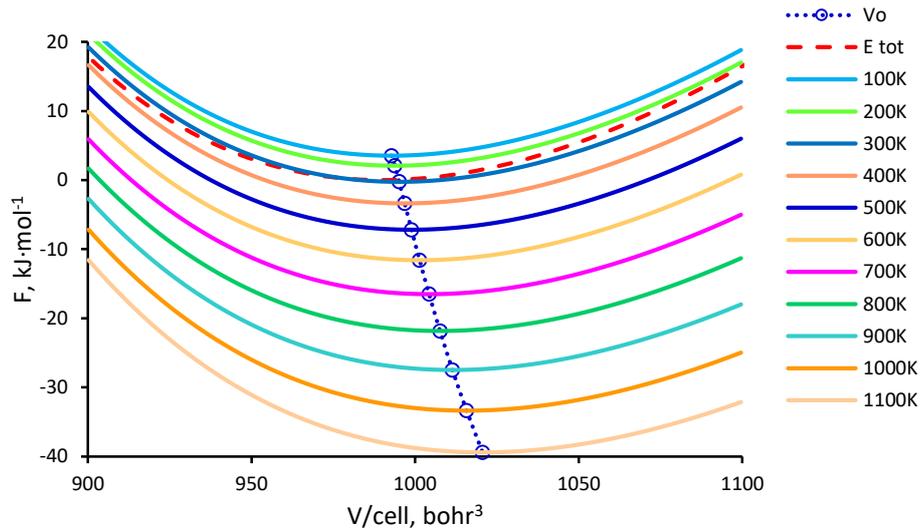

**Figure 5**. The free energy curves $F(V,T)$ of $Fe_2Mo$ calculated at different temperatures for the $n8$ path. The equilibrium volumes $V_0(T)$ are shown by the blue dotted line. The red dashed line represents the total energy $E_{tot}(V)$.

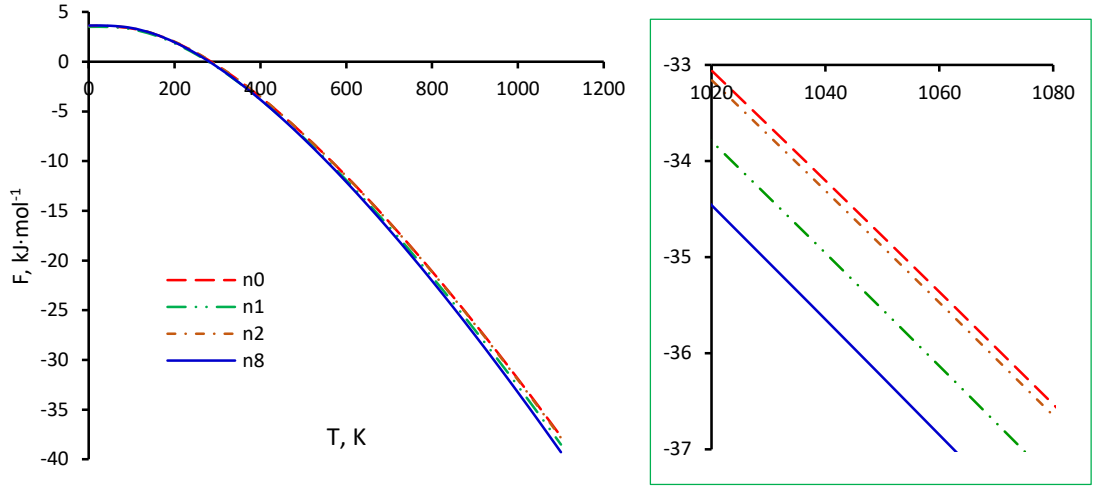

**Figure 6**. The free energy curves $F(T)$ of Fe$_2$Mo calculated along $n0$, $n1$, $n2$ and $n8$ paths, using first-principles phonon calculation. In the box on the right, the magnified values are given for convenience.

The free energies $F(V_0,T)$ calculated by (1) at corresponding equilibrium volumes along $n0$, $n1$, $n2$ and $n8$ paths are shown in Figure 6. According to this calculation the $n8$ possess the lowest energy among the other paths. Thus, $n8$ is the path of thermal expansion of Fe$_2$Mo.

The free energy of Fe$_2$Mo calculated by (1) using DFT-based phonon calculations obtained for the thermal expansion path $n8$ is shown in Figure 7 by the solid blue line. For the sake of comparison, the free energy of Fe$_2$Mo calculated using QDG approach and magnetic entropy [14] is shown in Figure 7 by the red dashed line.

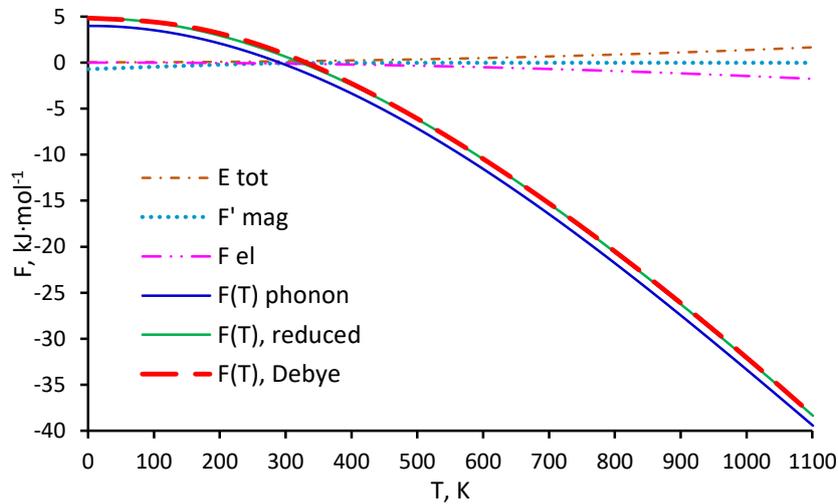

**Figure 7**. The free energy ($F(T)$, phonon) obtained as a sum of the total, electronic, magnetic and vibrational energy calculated using first-principles phonon calculation along the thermal expansion path of Fe$_2$Mo shown in comparison together with the free energy ($F(T)$, Debye) calculated using quasi-harmonic Debye - Grüneisen approximation reported in [14] and the free energy ($F(T)$, reduced) reduced to the zero-point energy expressed within the Debye model.

The total energy, electronic (2) and magnetic (7) energies contributing to the free energy of $Fe_2Mo$, are also shown in Figure 7.

Since the Debye model assumes the presence of zero vibrations [15], so if we bring the free energy calculated using phonons to the energy calculated using the Debye model, then we get the reduced function shown in Figure 7 by the solid green line.

The results presented in Figure 7 show that the free energy calculated in [14] using QDG closely match with the reduced free energy obtained using DFT-based phonon approach implemented in this work.

The *n8* path of thermal expansion of $Fe_2Mo$ which is the most energetically favorable among the other paths as follow from the calculations shown in Figure 6, is completely coincides with the path calculated in [14] using QDG approach and magnetic entropy.

*3.2.2. Thermodynamic properties of $Fe_2Mo$*

Thermodynamic properties were calculated along the most energetically stable thermal expansion path of $Fe_2Mo$. The calculated volume expansion *V(T)* of $Fe_2Mo$ is shown in Figure 8 (a) together with experimental data [10, 12, 13] and the theoretical calculation [14] for comparison.

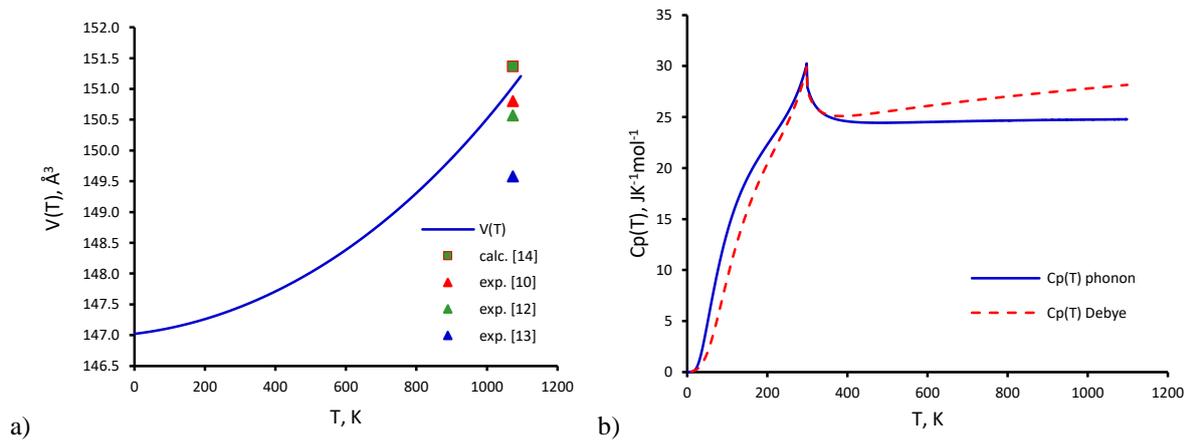

**Figure 8**. a) The volume expansion *V(T)* (in $Å^3$) calculated using first-principles phonon calculation along the thermal expansion path of $Fe_2Mo$ Laves phase in comparison with the experimental data [10, 12, 13] shown by the red, green and blue triangles; and the lattice parameters of $Fe_2Mo$ calculated in [14] using the quasi-harmonic Debye - Grüneisen approach is shown by the green square. b) The isobaric heat capacities $C_p(T)$ obtained using the first-principles based phonon approach and the Debye - Grüneisen model calculated along the thermal expansion path of $Fe_2Mo$. The kink is a ferro/paramagnetic phase transition predicted by used the model [24].

The isobaric heat capacity *Cp(T)* calculated along the thermal expansion path of $Fe_2Mo$ using DFT-based phonon and QDG approaches are plotted in Figure 8 (b) by the blue solid line and the red dashed line, a kink at 299 K [14] corresponds to the ferro/paramagnetic phase transition as the applied model [24] predicts.

As far as is known, there aren't any available experimental data on the ferro/paramagnetic phase transition temperature for $Fe_2Mo$, so a study of an accurate value of the Curie temperature is still necessary.

*3.2.3. Discussion*

In order to predict a trajectory of the thermal expansion path of Fe$_2$Mo the free energies were calculated along different directions intersected at the point ($a_0$, $b_0$), which is associated with T = 0 K. The selected four paths *n0*, *n1*, *n2* and *n8* are shown in Figure 2. The phonon dispersion curves and their correspondent density of states calculated along the *n8* path at the *pD*, ($a_0$, $c_0$), *pB* and *pE* points are shown in Figure 3 (d), Figure 3 (f), Figure 3 (b) and Figure 3 (e), respectively.

This order reflects the change in vibration energy from compression, which is at point *pD*, to expansion, which is at point *pE*, which corresponds to T = 1291 K, as predicted by the current calculation. At this temperature Fe$_2$Mo becomes unstable according to the Fe – Mo phase diagram estimated by Massalski et al. [28]. According to the diagram, at 1200 K in equilibrium, this compound undergoes a phase transformation and decomposes into a body-centered cubic (bcc) solid solution and a µ- phase. Thus, the imaginary phonon modes shown in Figure 3 (e), confirm that at this temperature the Fe$_2$Mo become an unstable compound indeed.

These figures reveal that with an increase in temperature during expansion along the *n8* path, the vibrational frequencies slightly decrease. This relatively slight frequencies softening with temperature implies that the Fe$_2$Mo can possess heat-resistant properties.

An analysis of the isotropic expansion along the *n0* path shown in Figure 3 (c), Figure 3 (f) and Figure 3 (a) reveals that with the increasing volume, the vibrational frequencies change insignificantly. Apparently, it can reflect the fact that this direction of thermal expansion is not an energetically favorable.

The graph of volume expansion of Fe$_2$Mo shown in Figure 8 (a) reveals that the values of lattice parameters calculated in this work at 1073 K using the DFT-based phonon calculation, are close to the experimental data [10] than the values reported in [14] obtained using QDG approach.

The heat capacity curves calculated using DFT-based phonon and QDG approaches are presented in Figure 8 (b). Both these curves possess a similar attitude of a kink due the same used model [24] describing the ferro/paramagnetic phase transition at T$_C$. But, at elevated temperatures the curve of heat capacity calculated using QDG approach continues to increase, while the curve calculated using DFT-based phonon approach goes to a horizontal straight line, according to Dulong–Petit law. According to calculations, the former curve will reach a horizontal line at higher temperatures.

The *n8* is the calculated path of thermal expansion of Fe$_2$Mo which passing near the experimental values of lattice parameters reported in [10, 12, 13], as shown in Figure 2. The comparison of the free energy calculated along the *n8* with the energies obtained for other paths calculated in this work using DFT-based phonon calculations and with the energies calculated along the same *ni* paths using QDG approach reported in [14], can bring us to the conclusion that these two methods used for STEP are equal to each other.

The use of the Debye – Grüneisen method makes it possible to estimate the influence of the vibration energy and the magnetic entropy of local magnetic moments of atoms on the stability of compounds. Whereas in the DFT-based phonon method, the magnetic contribution is already included into the calculations. Thus, it is difficult to evaluate them separately.

This calculation shows that the vibrational energy calculated using phonons differs from the vibrational energy obtained using QDG method, and therefore, if it is used in calculations, taking into account the magnetic entropy is a necessary condition for a correct description of the thermodynamics of compounds containing elements with magnetic properties.

The calculated path of thermal expansion of Fe$_2$Mo shows that at heating the lattice parameter *a*- will increase its value while the parameter *c*- decrease, and vice versa as the temperature decreases. In both cases, this will cause stresses in the matrix. Therefore, understanding the direction of the thermal expansion path of a compound can help to understand a strengthening mechanism of alloys and in the task of designing new nuclear reactors materials.

## 4. Conclusions

The first-principles phonon calculations have been carried out to calculate the vibrational energy contribution to the Helmholtz free energy of $Fe_2Mo$ Laves phases. The energetically favourable thermal expansion path of $Fe_2Mo$ was obtained by comparison the free energies calculated along different trajectories of thermal expansion. The obtained results show that the trajectory of thermal expansion path of $Fe_2Mo$ calculated using the quasi-harmonic Debye – Grüneisen approximation closely matches with the path obtained using DFT-based phonon method implemented in this work. Therefore, both of these approaches can equally be used for searching a thermal expansion path. The lattice volume expansion and the isobaric heat capacity were calculated. This study predicts that $Fe_2Mo$ possess a non-isotropic thermal expansion. The results of this work can be used to construct Gibbs potentials to refine the boundaries of the Fe-Mo phase diagram and further design fuel claddings for future nuclear reactors.

**Acknowledgments**

The research was financially supported by the Russian Foundation for Basic Research as a part of scientific project № 19-03-00530.